\documentstyle[array,epsfig,twocolumn,pra,aps]{revtex}

\begin{document}
\draft

\wideabs{

\title{Quantum information processing for a coherent superposition state via a mixed entangled coherent channel}

\author{H. Jeong,$^1$ M. S.
  Kim,$^1$ and Jinhyoung
  Lee$^{1,2}$}

\address{$^1$School of Mathematics and Physics, Queen's University,
Belfast BT7 1NN, United Kingdom \\
$^2$ Department of Physics, Sogang University, CPO Box 1142, Seoul
100-611, Korea }

\date{\today}

\maketitle

\begin{abstract}
  An entangled two-mode coherent state is studied within the framework
  of $2\times 2$ dimensional Hilbert space. An entanglement
  concentration scheme based on joint Bell-state measurements is
  worked out. When the entangled coherent state is embedded in vacuum
  environment, its entanglement is degraded but not totally lost. It
  is found that the larger the initial coherent amplitude, the faster
  entanglement decreases. We investigate a scheme to teleport a
  coherent superposition state while considering a mixed quantum
  channel. We find that the decohered entangled coherent state may be
  useless for quantum teleportation as it gives the optimal fidelity
  of teleportation less than the classical limit 2/3.
\end{abstract}

  }

%\maketitle

\section{INTRODUCTION}
\label{sec:intro}
Quantum entanglement and its remarkable features make it possible to
realize quantum information processing including quantum
teleportation\cite{Bennett93}, cryptography\cite{Ekert91}, and quantum
computation\cite{Barenco95}. An entanglement of two systems in
coherent states allows tests of local realism \cite{Sanders92} and can
be used as a quantum entangled channel for quantum information
transfer. Proposals to entangle fields in two spatially separated
cavities exist \cite{Davidovich93}. Recently, entanglement of
non-orthogonal states called quasi-Bell states and teleportation using
them have been studied \cite{Enk00,Hirota01,Wang01}.

In quantum information processing, the entangled coherent state is
normally categorized into a two-mode-continuous-variable state.
However, there was a suggestion to implement a logical qubit encoding
by treating a coherent superposition state, a
single-mode-continuous-variable state, as a qubit in 2-dimensional
Hilbert space \cite{C}. In this paper, we study the entangled coherent
state within the framework of $2\times 2$ dimensional Hilbert space.
We assess the entanglement of the evolved state and how useful it can
be to transfer the quantum information when the entangled coherent
state decoheres in the vacuum.

We first construct an orthogonal Bell basis set from non-orthogonal
coherent states to reformulate the problem to 2$\times$2 dimensional
Hilbert space. We then investigate the Bell-state measurement scheme
that works perfectly in the large amplitude limit. The measurement
scheme composed of linear devices is proposed to use for entanglement
concentration \cite{Bose99} and quantum teleportation. The
teleportation scheme, in effect, re-illustrates van Enk and Hirota's
\cite{Enk00}.  When the quantum system is open to the outside world,
the initially prepared system decoheres and becomes mixed. Assuming
the vacuum environment, we find how an entangled coherent state loses
its initial entanglement as interacting with the environment. We use
the measure of entanglement \cite{Lee00} based on the partial
transposition condition of entanglement \cite{PT96}. We then consider
optimal quantum teleportation via the mixed quantum channel. We find
that even though the channel is always entangled under the influence
of the vacuum environment, it becomes useless for teleportation at
some point.

\section{Construction of Bell basis with entangled coherent states}
\label{sec:2}
It is possible to consider an entangled coherent state in $C^2 \otimes
C^2$ Hilbert space. It makes the problem simpler because two-qubit
entangled states have the simplest mathematical structure among
entangled states.

Let us consider two kinds of entangled coherent states which have
symmetry in phase space:
\begin{eqnarray}
\label{eq:ecs12}
  |C_{1}\rangle&=&\frac{1}{\sqrt{N}}(|\alpha\rangle|\alpha\rangle +
  e^{i \varphi} |-\alpha^*\rangle|-\alpha^*\rangle), \\ 
\label{eq:ecs34} 
  |C_{2}\rangle&=&\frac{1}{\sqrt{N^\prime}}(|\alpha\rangle|-\alpha^*\rangle
  + e^{i \varphi^\prime} |-\alpha^*\rangle|\alpha\rangle),
\end{eqnarray}
where $|\alpha\rangle$ and $|-\alpha^*\rangle$ are coherent states of
amplitudes $\alpha$ and $-\alpha^*$, $N$ and $N^{\prime}$ are
normalization factors, and $\varphi$ and $\varphi^\prime$ are relative
phase factors. It can be verified that any entangled coherent states
in the form of $(|\beta\rangle|\beta\rangle+ e^{i
  \varphi}|\gamma\rangle|\gamma\rangle)/\sqrt{\cal N}$ or
$(|\beta\rangle|\gamma\rangle+ e^{i
  \varphi^\prime}|\gamma\rangle|\beta\rangle)/\sqrt{{\cal N}'}$, where
$\beta$ and $\gamma$ are any complex amplitudes, can be converted
respectively to $|C_1\rangle$ or $|C_2\rangle$ by applying local
unitary operations \cite{Kim}. A set of $|C_1\rangle$ for
$\varphi=0,\pi$ and $|C_2\rangle$ for $\varphi^\prime=0,\pi$ was
studied as quasi-Bell states\cite{Hirota01} but the four quasi-Bell
states do not form a complete measurement set by themselves because
they do not satisfy orthogonality and completeness.

By Gram-Schmidt theorem, it is always possible to make orthonormal
bases in $N$-dimensional vector space from any $N$ linear independent
vectors. Suppose new orthonormal bases by superposing non-orthogonal
and linear independent two coherent states $|\alpha\rangle$ and
$|-\alpha^*\rangle$:
\begin{eqnarray}
\label{eq:basis1}
|\psi_+\rangle&=&\frac{1}{\sqrt{N_\theta}}(\cos\theta e^{-\frac{1}{2}i\phi}|\alpha\rangle
-\sin\theta e^{\frac{1}{2}i\phi} |-\alpha^*\rangle),~~ \\
\label{eq:basis2}
|\psi_-\rangle&=&\frac{1}{\sqrt{N_\theta}}(-\sin\theta e^{-\frac{1}{2}i\phi}
|\alpha\rangle+\cos\theta e^{\frac{1}{2}i\phi} |-\alpha^*\rangle),
\end{eqnarray}
where $N_\theta=\cos^2 2 \theta$ is a normalization factor and real parameters $\theta$ and $\phi$
are defined as
\begin{equation}
\sin 2\theta e^{-i \phi}=\langle\alpha|-\alpha^*\rangle=\exp\{-2\alpha_r^2+2 i\alpha_r\alpha_i\}
\end{equation}
with real $\alpha_r$ and imaginary $\alpha_i$ parts of $\alpha$.

We define four maximally entangled Bell states using the
orthonormal bases in (\ref{eq:basis1}) and (\ref{eq:basis2}):
\begin{eqnarray}
\label{eq:bell1}
|B_{1,2}\rangle= 
\frac{1}{\sqrt{2}}( |\psi_{+}\rangle|\psi_{+}\rangle
\pm |\psi_{-}\rangle|\psi_{-}\rangle ), \\ 
\label{eq:bell2}
|B_{3,4}\rangle=
\frac{1}{\sqrt{2}}( |\psi_{+}\rangle|\psi_{-}\rangle \pm
|\psi_{-}\rangle|\psi_{+}\rangle).
\end{eqnarray}
They can be represented by
$|\alpha\rangle$ and $|-\alpha^*\rangle$ as
\begin{eqnarray}
\label{eq:bellalpha1}
\nonumber
|B_{1}\rangle&=& \frac {1} {\sqrt{2} N_{\theta}} \bigg\{
  e^{- i \phi}|\alpha\rangle|\alpha\rangle
  +e^{ i \phi}|-\alpha^*\rangle|-\alpha^*\rangle \\ 
  & &~~~~~~~~~-\sin 2\theta\Big(|\alpha\rangle|-\alpha^*\rangle+|-\alpha^*\rangle|\alpha\rangle\Big)
  \bigg\},\\
\label{eq:bellalpha2}
|B_{2}\rangle&=& 
  \frac {1} {\sqrt{2 N_{\theta}}} \Big(
  e^{ -i \phi}|\alpha\rangle|\alpha\rangle
  -e^{ i \phi}|-\alpha^*\rangle|-\alpha^*\rangle \Big),\\
\label{eq:bellalpha3}
\nonumber
  |B_{3}\rangle&=& 
  \frac {1} {\sqrt{2} N_{\theta}} \bigg\{
  |\alpha\rangle|-\alpha^*\rangle 
  +|-\alpha^*\rangle|\alpha\rangle \\
  & & -\sin 2\theta \Big(e^{-i \phi}|\alpha\rangle|\alpha\rangle
  +e^{i \phi}|-\alpha^*\rangle|-\alpha^*\rangle \Big) \bigg\},\\
\label{eq:bellalpha4} 
|B_{4}\rangle&=& 
\frac {1} {\sqrt{2 N_{\theta}}} \Big(
|\alpha\rangle|-\alpha^*\rangle
-|-\alpha^*\rangle|\alpha\rangle \Big),
\end{eqnarray}
where we immediately recognize that $|B_2\rangle$ and $|B_4\rangle$
are in the form of entangled coherent states $|C_1\rangle$ and
$|C_2\rangle$ while $|B_{1}\rangle$ and $|B_{3}\rangle$ become so as
$|\alpha|\rightarrow\infty$.

Now we are ready to consider decoherence and teleportation with mixed
entangled coherent states.  For simplicity we assume $\phi=0$, {\it
  i.e.}, $\alpha$ is real, in the rest of the paper.

\section{Teleportation via a pure channel}
\label{sec:sec3}
There have been studies on the quantum teleportation of a coherent
superposition state via an entangled coherent channel $|B_2\rangle$
\cite{Enk00}. Here, we suggest a scheme for the same purpose with the
use of Bell bases (\ref{eq:bellalpha1}), (\ref{eq:bellalpha2}),
(\ref{eq:bellalpha3}), and (\ref{eq:bellalpha4}).  The scheme includes
direct realization of Bell-state measurements. We also show that the
Bell-state measurement method enables the entanglement concentration
of partially entangled coherent states.

\subsection{Teleportation and Bell-state measurement}
\label{subsec:tandb}
Suppose a coherent superposition state
\begin{equation}
\label{eq:css}
|\Psi\rangle=\frac{1}{\sqrt{M_-}}(|\sqrt{2}\alpha\rangle-|-\sqrt{2}\alpha\rangle),
\end{equation}
where $M_-$ is a normalization factor, is superposed on a vacuum
$|0\rangle$ by a lossless 50:50 beam. It can be shown that the output
state is $|B_4\rangle$. It is possible to generate a superposition of
the two coherent states $|\sqrt{2}\alpha\rangle$ and
$|-\sqrt{2}\alpha\rangle$, from a coherent state
$|\sqrt{2}\alpha\rangle$ propagating through a nonlinear medium
\cite{Yurke}.

Let us assume that Alice wants to teleport a coherent superposition
state
\begin{equation}
\label{eq:unknown1}
|\psi\rangle_a={\cal A}|\alpha\rangle_a+{\cal B}|-\alpha\rangle_a
\end{equation}
via the pure entangled coherent channel $|B_4\rangle_{bc}$, where the
amplitudes ${\cal A}$ and ${\cal B}$ are unknown. The state (\ref{eq:unknown1})
can be represented as
\begin{equation}
\label{eq:unknown2}
|\psi\rangle_a={\cal A}^{\prime}|\psi_+\rangle_a+{\cal B}^{\prime}|\psi_-\rangle_a 
\end{equation}
with ${\cal A}^\prime={\cal A}\cos\theta+{\cal B}\sin\theta$ and
${\cal B}^\prime={\cal A}\sin\theta+{\cal B}\cos\theta$.  After
sharing the quantum channel $|B_4\rangle_{bc}$, Alice performs a
Bell-state measurement on her part of the quantum channel and the
state (\ref{eq:unknown1}) and sends the outcome to Bob. Bob
accordingly chooses one of the unitary transformations
$\{i\sigma_y,\sigma_x,-\sigma_z,\openone\}$ to perform on his part of
the quantum channel. Here $\sigma$'s are Pauli operators and
$\openone$ is the identity operator and the correspondence between the
measurement outcomes and the unitary operations are $B_1\Rightarrow
i\sigma_y$, $B_2\Rightarrow\sigma_x$, $B_3\Rightarrow -\sigma_z$,
$B_4\Rightarrow \openone$. Acting of these operators on
$|\alpha\rangle$ and$|-\alpha\rangle$ gives impacts as follows:
\begin{eqnarray}
\label{eq:op1}
&|&\alpha\rangle\stackrel{i\sigma_y}{\longrightarrow} \frac{1}{N_{\theta}}\big(\sin
2\theta|\alpha\rangle -|-\alpha\rangle\big), \\ 
  &|&-\alpha\rangle\stackrel{i\sigma_y}{\longrightarrow}
  \frac{1}{N_{\theta}}\big(|\alpha\rangle -\sin
  2\theta|-\alpha\rangle\big), \\ 
  &|&\alpha\rangle\stackrel{\sigma_x}{\longleftrightarrow}|-\alpha\rangle,
  \\ 
  &|&\alpha\rangle\stackrel{-\sigma_z}{\longrightarrow} \frac{1}{N_{\theta}}\big(|\alpha\rangle
  -\sin 2\theta|-\alpha\rangle\big), \\
\label{eq:op5} &|&-\alpha\rangle\stackrel{-\sigma_z}{\longrightarrow}
\frac{1}{N_{\theta}}\big(\sin 2\theta|\alpha\rangle
-|-\alpha\rangle\big).
\end{eqnarray}

It is not a trivial problem to discriminate all four Bell states. In
fact it was shown that complete Bell-state measurements on a product
Hilbert space of two two-level systems are not possible using linear
elements \cite{L}. We here suggest an experimental setup as shown in
Fig.~1 to discriminate Bell states constructed from entangled coherent
states.  Although perfect discrimination is not possible, arbitrarily
high precision can be achieved when the amplitude of the coherent
states becomes large.  For simplicity, we shall assume that the 50:50
beam splitter imparts equal phase shifts to reflected and transmitted
fields.

\begin{figure}
\label{fig:setup}
\centerline{\scalebox{.5}{\includegraphics{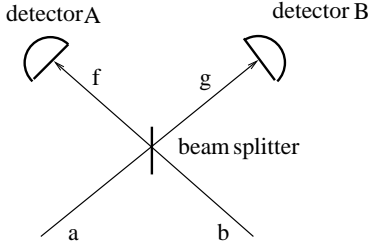}}} \vspace{0.3cm}
\caption{Scheme to discriminate all four Bell states with an
  arbitrarily high precision using a 50:50 beam splitter and two
  photo-detectors. If an odd number of photons is detected at detector
  A for mode $f$ then we know that the entangled state incident on the
  measurement set up was $|B_2\rangle$. On the other hand, if an odd
  number of photons is detected at detector B for mode $g$ then the
  incident entangled state was $|B_4\rangle$. For $\alpha\gg 1$, if a
  non-zero even number of photons is detected for mode $f$, the
  incident state was $|B_1\rangle$ and if a non-zero even number is
  detected for mode $g$, it was $|B_3\rangle$. }
\end{figure}

Suppose that each mode of the entangled state is incident on the beam
splitter. After passing the beam splitter (bs), the Bell states become
\begin{eqnarray}
\label{fig:setup1}
|B_1\rangle_{ab} &\stackrel{bs}\longrightarrow& {1\over \sqrt{2}N_\theta}
  (|even\rangle_f|0\rangle_g
  -\sin 2\theta|0\rangle_f|even\rangle_g),
  \nonumber \\
  |B_2\rangle_{ab} &\stackrel{bs}\longrightarrow & {1\over \sqrt{2N_\theta}}
  |odd\rangle_f|0\rangle_g,
  \nonumber \\
  |B_3\rangle_{ab} &\stackrel{bs}\longrightarrow & {1\over \sqrt{2}N_\theta}
  (|0\rangle_f|even\rangle_g, 
  -\sin 2\theta|even\rangle_f|0\rangle_g),
  \nonumber \\
  |B_4\rangle_{ab} &\stackrel{bs}\longrightarrow & {1\over \sqrt{2N_\theta}}
  |0\rangle_f|odd\rangle_g,
\end{eqnarray}
where $|even\rangle=|\sqrt{2}\alpha\rangle +|-\sqrt{2}\alpha\rangle$
has non-zero photon-number probabilities only for even numbers of
photons and $|odd\rangle=|\sqrt{2}\alpha\rangle
-|-\sqrt{2}\alpha\rangle$ has non-zero photon-number probabilities
only for odd numbers of photons. Note that $|even\rangle$ and
$|odd\rangle$ are not normalized. If an odd number of photons is
detected at detector A for mode $f$ then we know that the entangled
state incident on the measurement set up was $|B_2\rangle$. On the
other hand, if an odd number of photons is detected at detector B for
mode $g$ then the incident entangled state was $|B_4\rangle$. When
even numbers of photons are measured, we cannot in general tell if the
incident state was $|B_1\rangle$ or $|B_3\rangle$. However, for $\sin
2 \theta ~(=\langle\alpha|-\alpha\rangle) \simeq 0$, {\it i.e.}
$\alpha\gg 1$, if a non-zero even number of photons is detected for
mode $f$, the incident state was $|B_1\rangle$, and if a non-zero even
number is detected for mode $g$, it was $|B_3\rangle$. When $\sin
2\theta$ is not negligible, the probability of wrong estimation is
\begin{equation}
P_i(\alpha)=
\frac { 1 }
{ 2(1+e^{4\alpha^2}) }.
\end{equation}
For the limit of $\alpha \gg 1$, this probability approaches to zero
and all the Bell states can be discriminated with arbitrarily high
precision.

When the measurement outcome is $|B_2\rangle$, the receiver performs
$|\alpha\rangle\leftrightarrow|-\alpha\rangle$ on $c$. Such the phase
shift by $\pi$ can be done using a phase shifter whose action is
described by $R(\varphi)=\mbox{e}^{i\varphi a^\dag a}$:
\begin{equation}
R(\varphi) a R^\dag (\varphi) = a\mbox{e}^{-i\varphi},
\end{equation}
where $a$ and $a^\dagger$ are the annihilation and creation operators.
When the measurement outcome is $|B_4\rangle$, the receiver does
nothing on $c$ as the required unitary transformation is only the
identity operation $\openone$. When the outcome is $|B_3\rangle$, an
operator
$\frac{1}{N_\theta}(|\alpha\rangle\langle\alpha|-|-\alpha\rangle\langle-\alpha|)$
plays the corresponding role, which becomes a unitary operator for
$\alpha\gg 1$. When the outcome is $|B_1\rangle$, $\sigma_x$ and
$\sigma_z$ should be successively applied.

\subsection{Concentration of partial entanglement via entanglement swapping}
\label{subsec:cons}
If the initially prepared quantum channel is in a pure but not maximally entangled
state, the channel may be distilled to a maximally entangled state before 
using it for quantum information processing including teleportation. This process
is known as the entanglement concentration protocol
\cite{Bennett96b,BBPS96}. For an entangled coherent channel, it can be
simply realized via entanglement swapping \cite{Bose99,Z} using the
Bell measurement proposed in Sec.~\ref{subsec:tandb}.

Suppose an ensemble of a partially entangled pure state
\begin{equation}
\label{D-4-extra}
|D_4\rangle=\frac{1}{\sqrt{N_\eta}}
(\cos\eta|\alpha\rangle|-\alpha\rangle-\sin\eta|-\alpha\rangle|\alpha\rangle )
\end{equation}
from which we want to distill a sub-ensemble of a maximally entangled
state. $N_\eta$ is a normalization factor and the real phase factor
$\eta$, $0<\eta<\pi/2$, determines the degree of entanglement for
$|D_4\rangle$. The state $|D_4\rangle$ in (\ref{D-4-extra}) is written in
the orthonormal bases (\ref{eq:basis1}) and (\ref{eq:basis2}) as follows:
\begin{eqnarray}
|D_4\rangle=\frac{1}{\sqrt{N_\eta}} \bigg\{&&\frac{1}{2}\sin
2\theta(\cos\eta-\sin\eta)\big(|\psi_+\rangle|\psi_+\rangle+|\psi_-\rangle|\psi_-\rangle\big)\nonumber
\\
&&+\big(\cos^2\theta\cos\eta-\sin^2\theta\sin\eta\big)|\psi_+\rangle|\psi_-\rangle\nonumber\\
\label{eq:D4-2}
&&+\big(\sin^2\theta\cos\eta-\cos^2\theta\sin\eta\big)|\psi_-\rangle|\psi_+\rangle    \bigg\}.
\end{eqnarray}

First, we consider the case when $\alpha$ is large.  In this case, state $|D_4\rangle\simeq
|E_4\rangle$ where
\begin{equation}
  \label{eq:1}
  |E_4\rangle= 
  \cos\eta|\psi_{+}\rangle|\psi_{-}\rangle -
  \sin\eta|\psi_{-}\rangle|\psi_{+}\rangle.
\end{equation}
After sharing a quantum channel between Alice and Bob, Alice prepares
a pair of particles which are in the same entangled state as the quantum
channel. Alice then performs Bell-state measurement on her pair of the
quantum channel.  If the measurement outcome is $B_1$ or $B_2$, the
other particle of Alice's and Bob's quantum channel is, respectively,
in maximally entangled state $|B_1\rangle_{b^\prime c}$ or $
|B_2\rangle_{b^\prime c}$ where Alice's particle is denoted by
$b^\prime$. Otherwise, Alice's particle and Bob's quantum channel are
not in a maximally entangled state:
\begin{eqnarray}
\label{eq:33}
|B_3^\prime\rangle_{b^\prime c}&=& \frac{1}{\sqrt{N_\eta^\prime}}(
\cos^2\eta|\psi_{+}\rangle_{b^\prime}|\psi_{-}\rangle_c +
  \sin^2\eta|\psi_{-}\rangle_{b^\prime}|\psi_{+}\rangle_c ), \\
\label{eq:44}
|B_4^\prime\rangle_{b^\prime c}&=& \frac{1}{\sqrt{N_\eta^\prime}}(
\cos^2\eta|\psi_{+}\rangle_{b^\prime}|\psi_{-}\rangle_c -
\sin^2\eta|\psi_{-}\rangle_{b^\prime}|\psi_{+}\rangle_c ),
\end{eqnarray}
respectively for measurement outcome of $B_3$ or $B_4$.
$N_\eta^\prime$ is a normalization factor.  The probability $P_{1}$
and $P_2$ to obtain the maximally entangled state
$|B_1\rangle_{b^\prime c}$ and $|B_2\rangle_{b^\prime c}$ are
$P_1=P_2= \cos^2\eta \sin^2\eta$.  In this way, no matter how small
the initial entanglement is, it is possible to distill some maximally
entangled coherent channels from partially entangled pure channels.

We now consider the concentration protocol when $\alpha$ is not large
enough to neglect $\sin 2\theta$.  In this case, only two Bell-states
$|B_2\rangle$ and $|B_4\rangle$ can be precisely measured.  Extending
the previous argument leading to (\ref{eq:44}), when the measurement
outcome is $B_4$, the resulting state for particles $b^\prime$ and $c$
is not maximally entangled.  However, we can find that, for the
measurement outcome of $B_2$, the resulting state is
$|B_2\rangle_{b^\prime c}$ even for the case of $\alpha$ small. The
success probability ${\cal P}_2$ for this case is
\begin{equation}
{\cal P}_2(\theta,\eta)=\frac{\cos^4 2\theta \sin^2 2\eta}{4(1-\sin^2
  2\theta \sin 2\eta)}
\end{equation}
where ${\cal P}_2\rightarrow0$ for $\alpha\simeq0$ and ${\cal
  P}_2\rightarrow\cos^2\eta\sin^2\eta$ for $\alpha\gg1$.

\section{Decay of the entangled coherent channel: measure of entanglement}
\label{sec:sec4}
When the entangled coherent channel $|B_4\rangle$ is embedded in a
vacuum environment, the channel decoheres and becomes a mixed state of
its density operator $\rho_4(\tau)$, where $\tau$ stands for the
decoherence time. To know the time dependence of $\rho_4(\tau)$, we
have to solve the master equation \cite{Phoenix}
\begin{equation}
\label{master-eq}
  {\partial \rho \over \partial \tau}=\hat{J}\rho +\hat{L}\rho~;~\hat{J}\rho=\gamma a\rho a^\dag,~~
  \hat{L}\rho=-{\gamma \over 2}(a^\dag a\rho +\rho a^\dag a)
\end{equation}
where $\gamma$ is the energy decay rate. The formal solution of the
master equation (\ref{master-eq}) can be written as
\begin{equation}
\label{formal-sol}
  \rho(t)=\exp[(\hat{J}+\hat{L})\tau]\rho(0).
\end{equation}
which leads to the solution for the initial $|\alpha\rangle\langle\beta|$
\begin{equation}
\label{solution-master}
  \exp[(\hat{J}+\hat{L})\tau]|\alpha\rangle\langle\beta|=\langle\beta|\alpha\rangle^{1-t^2}
  |\alpha t\rangle\langle\beta t|
\end{equation}
where $t=e^{-\frac{1}{2}\gamma\tau}$. For later use, we introduce a
normalized interaction time $r$ which is related to $t$:
$r=\sqrt{1-t^2}$.

To restrict our discussion in a $2\times 2$ dimensional Hilbert space
even for the mixed case, the orthonormal basis vectors
(\ref{eq:basis1}) and (\ref{eq:basis2}) are now $\tau$-dependent:
%-> becomes now
\begin{eqnarray}
\label{eq:decaybasis1}
  |\Psi_+(\tau)\rangle&=&\frac{1}{\sqrt{N_\Theta}}(\cos\Theta |t\alpha\rangle
  -\sin\Theta |-t\alpha\rangle),~~ \\
  \label{eq:decaybasis2}
  |\Psi_-(\tau)\rangle&=&\frac{1}{\sqrt{N_\Theta}}(-\sin\Theta
  |t\alpha\rangle+\cos\Theta |-t\alpha\rangle),
\end{eqnarray}
where $\sin 2\Theta =\exp(-2 t^2\alpha^2)$. The unknown state to
teleport and the Bell-state bases are newly defined according to the
new basis vectors Eqs.~(\ref{eq:decaybasis1}) and
(\ref{eq:decaybasis2}).

Any two dimensional bipartite state can be written as
\begin{equation}
  \rho=\frac{1}{4}\bigg( \openone\otimes \openone+ {\vec v} \cdot {\vec
  \sigma}\otimes \openone + \openone \otimes {\vec s} \cdot {\vec \sigma} +
  \sum_{m,n=1}^3 t_{nm} \sigma_n\otimes\sigma_m \bigg),
\end{equation}
where coefficients $t_{nm}={\mathrm Tr}(\rho\sigma_m\otimes\sigma_n)$
form a real matrix $T$. Vectors $\vec v$ and $\vec s$ are local
parameters which determine the reduced density operator of each mode
\begin{eqnarray}
  \rho_b&=&{\mathrm Tr}_c\rho=\frac{1}{2}(\openone+ {\vec v}\cdot {\vec
  \sigma}),\\
  \rho_c&=&{\mathrm Tr}_b\rho=\frac{1}{2}(\openone+ {\vec s}\cdot {\vec
  \sigma}),
\end{eqnarray}
while the matrix $T$ is responsible for correlation\cite{Horodecki96}
\begin{equation}
  {\cal E}(a,b)={\mathrm Tr}(\rho\, {\vec a\cdot\vec \sigma\otimes
  \vec b\cdot\vec \sigma})=({\vec
  a},T {\vec b}).
\end{equation}
With use of Eqs.~(\ref{eq:bellalpha4}) and (\ref{solution-master}), we
find $\vec v$, $\vec s$, and $T$ for the mixed channel $\rho_4(\tau)$
as follows
\begin{eqnarray}
\label{eq:v}
  {\vec v}&=&{\vec s}=\bigg(\frac{B}{N_\theta},~0,~0\bigg), \\
  \label{eq:T}
  T&=&\frac{1}{2 N_\theta}\left(
  \begin{array}{ccc}
  A+D & 0 & 0 \\
  0 & -A+D & 0 \\
  0 & 0 & A-C 
\end{array}\right) \label{eq:Tmatrix},
\end{eqnarray}
where $A$, $B$, $C$ and $D$ are defined as
\begin{eqnarray}
  A&=& (1-\Gamma)\exp(-4 t^2 \alpha^2), \nonumber \\
  B&=& (1-\Gamma)\exp(-2 t^2 \alpha^2), \nonumber \\
  C&=& 2-(1+\Gamma)\exp(-4 t^2 \alpha^2), \nonumber \\
  D&=& -2\Gamma+(1+\Gamma)\exp(-4 t^2 \alpha^2), \nonumber \\
  \Gamma&=& \exp[-4(1-t^2)\alpha^2].
\end{eqnarray}
Note that $N_\theta$ is a time-independent
normalization factor and  $\rho_4(\tau\neq 0)$ can not be represented by a
Bell-diagonal matrix.

The necessary and sufficient condition for separability of a two
dimensional bipartite system is the positivity of the partial
transposition of its density matrix \cite{PT96}. Consider a density
matrix $\rho$ for a 2$\times$2 system and its partial transposition
$\rho^{T_2}$. The density matrix $\rho$ is inseparable iff
$\rho^{T_2}$ has any negative eigenvalue(s). We define the measure of
entanglement $E$ for $\rho$ in terms of the negative eigenvalues
of $\rho^{T_2}$ \cite{Lee00}. The measure of entanglement $E$ is
then defined as
\begin{equation}
\label{eq:moet}
  E = - 2\sum_i \lambda^-_i
\end{equation}
where $\lambda^-_i$ are the negative eigenvalue(s) of $\rho^{T_2}$ and
the factor 2 is introduced to have $0 \le E \le 1$.

For $\rho_4(\tau)$ we find the time evolution of the measure of
entanglement
\begin{equation}
\label{eq:moe}
  E(\tau)=\frac{\sqrt{16 B^2 + (C-D)^2}-(2A+C+D)} {4 N_\theta}.
\end{equation}
Initially, the state $|B_4\rangle$ is maximally entangled, {\it i.e.},
$E(\tau=0)=1$, regardless of $\alpha$. It is seen in Fig.~2(a) that
the mixed state $\rho_4(\tau)$ is never separable at the interaction
time $\tau <\infty$. It should be noted that the larger the initial
amplitude $\alpha$, the more rapidly the entanglement is degraded. It
is known that the speed of destruction of quantum interference depends
on the distance between the coherent component states\cite{Kim92}.
When the amplitudes of coherent component states are larger, the
entanglement due to their quantum interference is more fragile.

\vspace{1cm}

\begin{figure}
\label{fig:e-f}
\centerline{\scalebox{.7}{
\includegraphics{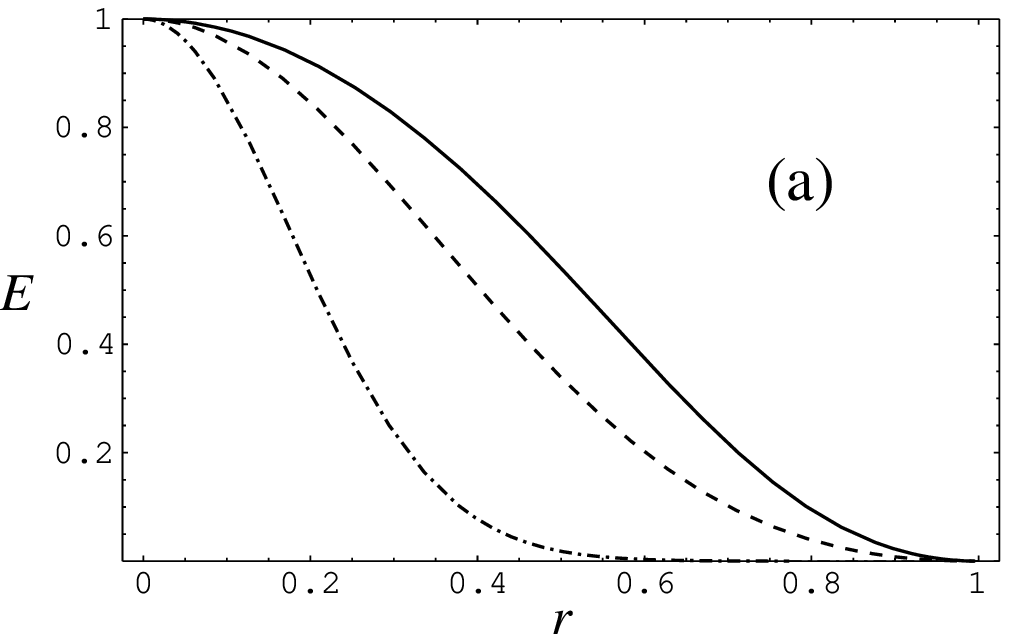}}}\vspace{0.4cm}
\centerline{\scalebox{.7}{
\includegraphics{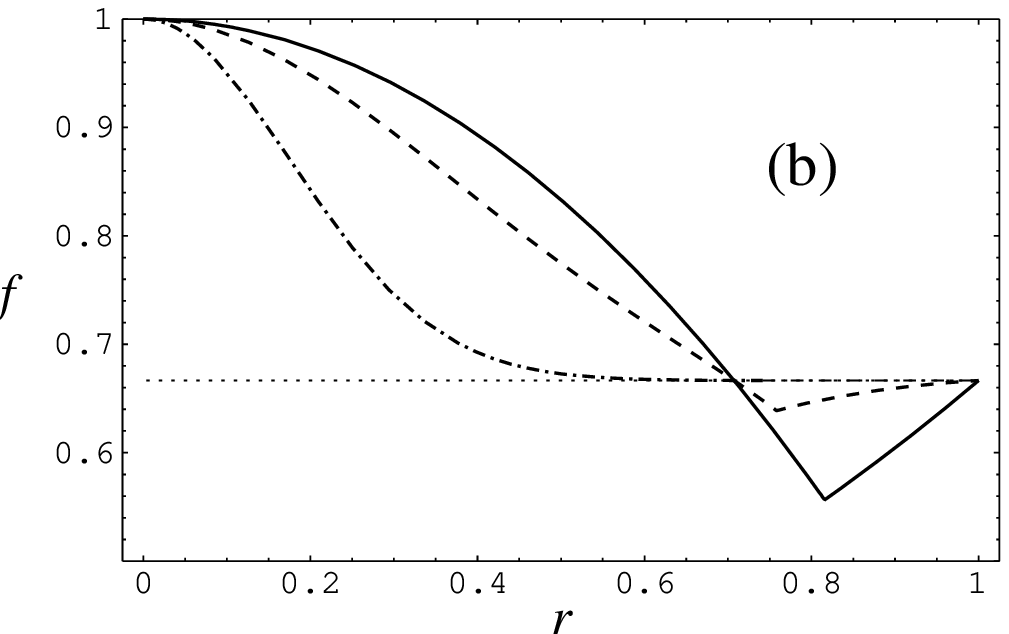}}}
 \vspace{0.2cm}
\caption{(a) Entanglement $E$ for the
  mixed entangled coherent channel against the normalized decoherence
  time $r=\sqrt{1-e^{-\gamma \tau}}$.  (b) Optimal fidelity $f$ of
  quantum teleportation with the mixed entangled coherent channel. The
  maximum fidelity $2/3$ obtained by classical teleportation is
  plotted by a dotted line. We can clearly see that the mixed channel
  is not useful in quantum teleportation from $r=1/\sqrt{2}$ even
  though it is always entangled. $\alpha=0.1$ (solid line), $\alpha=1$
  (long dashed) and $\alpha=2$ (dot dashed). }
\end{figure}

\section{Teleportation via a mixed channel}
\label{sec:sec5}

The optimal fidelity of teleportation in any general scheme by means
of trace-preserving local quantum operations and classical
communication (LQCC) via a single channel can be obtained from the
maximal singlet fraction of the channel\cite{Horodecki99}. The
relation is
\begin{equation}
\label{eq:fmax}
  f(\rho)=\frac{F(\rho) N+1}{N+1},
\end{equation}
where $f(\rho)$ is the optimal fidelity for the given quantum channel
$\rho$, $F(\rho)$ is the maximal singlet fraction of the channel, and
$N$ is the dimension of the related Hilbert space $C^N\otimes C^N$.
$F(\rho)$ is defined as max$\langle\Phi|\rho|\Phi\rangle$ where the
maximum is taken over all the $N\times N$ maximally entangled states.

Any $2\times 2$ channel becomes useless for quantum teleportation when
the optimal fidelity $f(\rho)$ is less than the classical limit $2/3$.
In other words, when $F(\rho)\leq 1/2$, the channel is useless for
quantum teleportation. To find the maximally entangled basis in which
a given channel has the highest fraction of a maximally entangled
state, it suffices to find rotations which diagonalize $T$
\cite{Horodecki97}. In the case of $\rho_4$, $T$ in (\ref{eq:Tmatrix})
is always a diagonal matrix. This means that the Bell
bases constructed from Eqs.
(\ref{eq:decaybasis1}) and (\ref{eq:decaybasis2}) give the maximal
singlet fraction at any decay time. The optimal fidelity $f(\rho_4)$
obtained by Eq.~(\ref{eq:fmax}) and the definition of the maximal
singlet fraction is
\begin{eqnarray}
\label{eq:fmax2}
  f(\rho_4)=\frac{1}{3}{\rm max}\bigg\{&1&
  +\frac{e^{4 \alpha^2 } - e^{4 t^2\alpha^2}}{e^{4\alpha^2}-1}, \nonumber \\
  &&\frac{e^{4 t^2\alpha^2}-e^{4 r^2\alpha^2}
  +2 e^{4\alpha^2}-2}{e^{4\alpha^2}-1}
  \bigg\}.
\end{eqnarray}
Because the initially defined Bell bases always give the maximal
singlet fraction, the optimal fidelity is obtained by the standard
teleportation scheme with Bell measurement and unitary operations.
This means that the experimental proposal in Sec.~\ref{sec:sec3} for
pure channel can also be used for a mixed channel to obtain the
optimal fidelity. The optimal fidelity for the standard teleportation
scheme is
\begin{equation}
\label{eq:fmaxst}
  f_{s}(\rho_4)={\rm max}\bigg[\frac{1}{2}(1-\frac{1}{3}{\mathrm Tr} TO)\bigg]=f(\rho_4),
\end{equation}
where the maximum is taken over all possible rotations
$O=O^+(3)$\cite{Horodecki96}. As the interaction time varies, parameters ${\vec v}$, ${\vec s}$
and $T$ are changed.  For the decoherence model we
consider in this paper, $T$ alone affects the fidelity of
teleportation.

Fig.~2(b) shows the optimal fidelity at the normalized
decay time $r$. The channel is always entangled as shown in
Fig.~2(a). However, after the characteristic time $r_c=1/\sqrt{2}$
the channel becomes useless for teleportation. It is worth noting that
the characteristic time does not depend on the initial $\alpha$ value.
This is confirmed by the fact that the only real solution of the
equation $f(\rho_4)=2/3$ is $r=1/\sqrt{2}$ regardless of $\alpha$.
Bennett {\em et al.} \cite{Bennett96} have pointed
out that some states with nonzero entanglement do not have the maximal
singlet fractions higher than $1/2$. %[ We have found that the ->]
The decohered entangled coherent channel $\rho_4(r\geq r_c)$ is 
an example of such the case.

Bose and Vedral \cite{Bose00} found that not only entanglement but
also mixedness of quantum channels affect the fidelity of
teleportation. We may conjecture that the higher entanglement and the
lower mixedness (higher purity) result in the better fidelity. In this
case, it is shown to be true only when the channel is useful for
teleportation. The mixedness of a given state $\rho$ can be quantified
by its linear entropy $S(\rho)=1-{\mathrm Tr}(\rho^2)$.  For the
decohered entangled coherent channel, the linear entropy is
\begin{equation}
  S(\rho_4)=
\frac{(e^{8r^2\alpha^2}-1)(e^{8t^2\alpha^2}-1)}     {2(e^{4\alpha^2}-1)^2},
\end{equation}
which increases to the maximal value and then decreases to zero as
shown in Fig.~3 because the channel interacts with the vacuum and the state
for $\tau\rightarrow\infty$ approaches to the two-mode vacuum which
is a pure state. We found that mixedness becomes maximized at the
characteristic time $r_c$. It is confirmed by solving the equation
${\partial S(\rho_4)}/{\partial r}=0$ which yields a unique real
solution $r=1/\sqrt{2}~=r_c$ again regardless of $\alpha$. It is
easily checked that von-Neumann entropy as a measurement of mixedness
gives exactly the same result.

\begin{figure}
\label{fig:mixedness}
\centerline{\scalebox{.7}{
\includegraphics{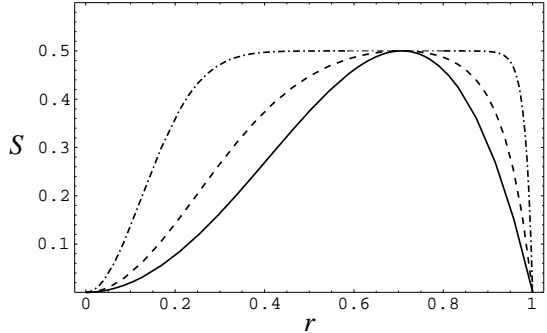}}}
 \vspace{0.2cm}
\caption{Mixedness $S$ quantified by the linear entropy for the mixed
  entangled coherent state against the normalized decoherence time
  $r$. The mixedness becomes maximized at the characteristic time
  $r_c$ after which the channel is no longer useful for teleportation.
  $\alpha=0.1$ (solid line), $\alpha=1$ (long dashed) and $\alpha=2$
  (dot dashed). }
\end{figure}

Horodecki {\it et al.} \cite{Horodecki97} showed that any entangled
$2\times 2$ density matrix can be distilled to a singlet form by local
filtering \cite{BBPS96,Gisin96} and entanglement concentration
protocol \cite{Bennett96b}. If sufficiently many entangled $2\times2$
channels are given, no matter how small the entanglement of the
channels is, some maximally entangled channels can be obtained from
the original pairs. Because the decohered channel $\rho_4$ is
entangled at any decay time, the ensemble represented by
$\rho_4(\tau)$ can be purified to obtain some maximally entangled
channels. We have seen that the singlet fraction $F(\rho_4)$ becomes
smaller than 1/2 after $r_c$, meanwhile the purification protocol in
\cite{Bennett96b} can be applied when the singlet fraction of a given
density matrix is larger than 1/2. Therefore, if the decay time is
longer than $r_c$, a local filtering or a
generalized measurement \cite{Horodecki97} should be first performed
on $\rho_4$ for purification. It has been pointed out that the
filtering process allows one to transfer the entanglement hidden in
the relation between $\vec v$, $\vec s$ and $T$ (the entanglement
added by change of the local states) to $T$ \cite{Horodecki97}.

\section{Usefulness for continuous-variable teleportation}
\label{sec:sec6}
We have studied entangled coherent states in $2\times 2$ Hilbert
space. However, entangled coherent states are in fact
continuous-variable states in infinite dimensional Hilbert space. If
$|B_2\rangle$ and $|B_4\rangle$ are considered in infinite dimensional
Hilbert space, they are not maximally entangled any
more\cite{Parker00}. It is thus natural to ask such a question: how
useful are the entangled coherent states for teleportation of
continuous-variable states?

In the protocol proposed in \cite{Braunstein98} and demonstrated
experimentally in \cite{Furusawa98} for continuous-variable
teleportation, a two-mode squeezed state is used as the quantum
channel and a joint homodyne measurement as Alice's measurement. An
unknown quantum state in Eq.~(\ref{eq:unknown1}) can be teleported by
a two-mode squeezed state, and the fidelity becomes unity for the
limit of infinite squeezing.

Assume that a coherent state of an unknown amplitude is the state to
teleport via an entangled coherent state, $|C_2\rangle$ in
Eq.~(\ref{eq:ecs34}) with $\varphi^\prime=0$. After a straightforward
calculation, the fidelity is obtained \cite{Lee00b}
\begin{equation}
\label{eq:fcont}
  f(\alpha)=\frac{1+\exp(-2\alpha_r^2)}{2\big[1+\exp(-4\alpha_r^2)\big]}.
\end{equation}
Note that $f(\alpha)$ is independent from the amplitude of the unknown
coherent state to teleport. It depends only on the real part of
coherent amplitude $\alpha$ of the quantum channel. We find from
Eq.~(\ref{eq:fcont}) that the fidelity is always better than 1/2. The
maximal value is about $0.6$ when $\alpha_r\simeq\pm0.7$.

\section{Remarks}
\label{sec:remarks}
We have studied a mixed entangled coherent channel in $2\times 2$
Hilbert space. We constructed orthogonal Bell bases with entangled
coherent states to consider their entanglement and usefulness for
teleportation in a dissipative environment. A pure entangled coherent
channel is shown to teleport perfectly some quantum information. We
investigated an experimental scheme for teleportation and entanglement
concentration with a realizable Bell-measurement method.

It is found that a mixed entangled coherent state is always entangled
regardless of the decay time. The larger initial amplitude $\alpha$,
the more rapidly entanglement is degraded. This is in agreement with
the fact that macroscopic quantum effects are not easily seen because
it is more fragile.

Because a decohered entangled coherent channel is entangled at any
decay time, its ensemble can be purified by an entanglement
purification protocol \cite{Bennett96b} and used for reliable
teleportation. On the other hand, it is shown that the optimal
fidelity of teleportation attainable using a single pair is better
than the classical limit $2/3$ only until a certain characteristic
time $r_c$, at which the mixedness of the channel becomes maximized.
The maximal singlet fraction of the state is not more than $1/2$ after
$r_c$, even though it is still entangled.

Entanglement and mixedness \cite{Bose00} of quantum channels are
important factors which affect teleportation. Until the characteristic
time, both entanglement and purity decrease, which causes the decrease
of teleportation fidelity. After the time $r_c$, the purity of the
channel is recovered back even though entanglement decreases further.
The experimental realization of purification for the mixed channels
deserves further investigation.

\acknowledgements

This work has been supported by the UK Engineering and Physical
Sciences Research Council (EPSRC) (GR/R 33304). JL thanks the Brain Korea 21
project (D-1099) of the Korea Ministry of Education for financial support.

%\onecolumn
%\newpage
%\twocolumn


\begin{references} 

\bibitem{Bennett93} C. H. Bennett, G. Brassard, C. Cr\'{e}peau, R.
  Jozsa, A. Peres, and W. K. Wootters, \prl {\bf 70}, 1895 (1993) ; D.
  Bouwmeester, J. W. Pan, K. Mattle, M. Eibl, H. Weinfurter, and A.
  Zeilinger, Nature {\bf 390}, 575 (1997).

\bibitem{Ekert91} A. Ekert,\prl {\bf 67}, 661 (1991).

\bibitem{Barenco95} A. Barenco, D. Dutch, A. Ekert, and R Jozsa, \prl
  {\bf 74}, 4083 (1995).

\bibitem{Sanders92} B. C. Sanders, \pra {\bf 45}, 6811 (1992); B. C.
  Sanders, K. S. Lee and M. S. Kim, \pra {\bf 52}, 735 (1995).

\bibitem{Davidovich93} L. Davidovich, A. Maali, M. Brune, J. M.
  Raimond, and S. Haroche, \prl {\bf 71}, 2360 (1993).

\bibitem{Enk00} S. J. van Enk and O. Hirota, e-print quant-ph/0012086.

\bibitem{Hirota01} O. Hirota and M. Sasaki, e-print quant-ph/0101018;
  O. Hirota, S. J. van Enk, K. Nakamura, M. Sohma, and K. Kato,
  e-print quant-ph/0101096.

\bibitem{Wang01} X. Wang, e-print quant-ph/0102011; X. Wang, e-print
  quant-ph/0102048.

\bibitem{C} P. T. Cochrane, G. J. Milburn, and W. J. Munro, \pra {\bf 59},
2631 (1999); W. J. Munro, G. J. Milburn and B. C. Sanders, \pra {\bf 62},
052108 (2000).

\bibitem{Bose99} S. Bose, V. Vedral, and P. L. Knight \pra {\bf 60},
  194 (1999).

\bibitem{Lee00} J. Lee and M. S. Kim, \prl {\bf 84}, 4236 (2000); J.
  Lee, M. S. Kim, Y.-J. Park, and S. Lee, J. Mod. Opt. {\bf 47}, 2151
  (2000).

\bibitem{PT96} A. Peres, \prl {\bf 76}, 1413 (1996); M. Horodecki, P.
  Horodecki, and R Horodecki, Phys. Lett. A {\bf 223}, 1 (1996).

\bibitem{Kim} D. Wilson, H. Jeong, M. S. Kim, and J. Lee, to be
  published.

\bibitem{Yurke} B. Yurke and D. Stoler, \prl {\bf 57}, 13 (1986).

\bibitem{L} N. L\"{u}tkenhaus, J. Calsamiglia, and K.-A. Suominen,
  \pra {\bf 59}, 3295 (1999).

\bibitem{Bennett96b} C. H. Bennett, S. Popescu, B. Schumacher, J. A.
  Smolin, and W. K. Wootters, \prl {\bf 76}, 722 (1996).

\bibitem{BBPS96} C. Bennett, H. J. Bernstein, S. Popescu, and B.
  Schumacher, \pra {\bf 53}, 2046 (1996).

\bibitem{Z} M. Zukowski, A. Zeilinger, M. A. Horne, and A. K. Ekert,
  \prl {\bf 71}, 4287 (1993).

\bibitem{Phoenix} S. J. D. Phoenix, \prl {\bf 41}, 5132 (1990).

\bibitem{Kim92} M. S. Kim and V. Buz\v{e}k, \pra {\bf 46}, 4239
  (1992); H. Jeong, J. Lee, M. S. Kim, \pra {\bf 61}, 052101 (2000).

\bibitem{Horodecki99} R. Horodecki, P. Horodecki, and M. Horodecki,
  \pra {\bf 60}, 1888 (1999).

\bibitem{Horodecki97} M. Horodecki, P. Horodecki, and R. Horodecki,
  \prl {\bf 78}, 574 (1997).

\bibitem{Horodecki96} R. Horodecki and M. Horodecki, and P. Horodecki.
  Phys. Lett. A {\bf 222}, 21 (1996); R. Horodecki and M. Horodecki,
  \pra {\bf 54}, 1838 (1996).

\bibitem{Bennett96} C. H. Bennett, D. P. DiVincenzo, J. A. Smolin, and
  W. K. Wootters, \pra {\bf 54}, 3824 (1996).

\bibitem{Bose00} S. Bose and V. Vedral, \pra {\bf 61}, 040101(R)
  (2000).

\bibitem{Gisin96} N. Gisin, Phys. Lett. A {\bf 210}, 151 (1996).

\bibitem{Parker00} S. Parker, S. Bose, M. B. Plenio \pra {\bf 61},
  32305 (2000).

\bibitem{Braunstein98} S. L. Braunstein and H. J. Kimble, \prl, {\bf
  80}, 869 (1998).

\bibitem{Furusawa98} A. Furusawa, J. L. S{\o}rensen, S. L. Braunstein,
  C. A. Fuchs, H. J. Kimble, and E. S. Polzik, Science {\bf 282}, 706
  (1998).

\bibitem{Lee00b} J. Lee, M. S. Kim, and H. Jeong, \pra {\bf 62},
  032305 (2000).


\end{references}
\end{document}